\documentclass[11pt,letterpaper]{article}

\pdfoutput=1 
\usepackage{color}
\usepackage{jheppub}
\usepackage{amsmath}
\usepackage{amsfonts}
\usepackage{amssymb}
\usepackage{graphicx}
\usepackage{epsfig}
\usepackage{natbib}

\usepackage{subcaption}
\usepackage[thinlines]{easytable}
\usepackage{array}
\usepackage[font=small]{caption}
\def\ba{\begin{eqnarray}}
\def\ea{\end{eqnarray}}
\def\bea{\begin{eqnarray}}
\def\eea{\end{eqnarray}}
\def\be{\begin{equation}}
\def\ee{\end{equation}}

\def\({\left(}
\def\){\right)}
\def\[{\left[}
\def\]{\right]}

\newcolumntype{M}[1]{>{\centering\arraybackslash}m{#1}}

\title{ Schr\"{o}dinger field theory in curved spacetime: In-In formalism and three-point function for inflationary background}
\author[a,b]{Abasalt Rostami,}

\affiliation[a]{Department of Physics, Sharif University of Technology,
	Tehran, Iran }
\affiliation[b]{ School of Physics, Institute for Research in Fundamental Sciences (IPM), P. O. Box 19395-5531, Tehran, Iran }

\author[c]{Javad T. Firouzjaee}
\affiliation[c]{ School of Astronomy, Institute for Research in Fundamental Sciences (IPM), P. O. Box 19395-5531, Tehran, Iran }
\emailAdd{aba-rostami@ipm.ir}
\emailAdd{j.taghizadeh.f@ipm.ir}

\abstract{	We review the Schr\"{o}dinger picture of field theory in curved spacetime and using this formalism, the power spectrum of massive non-interacting, minimally coupled scalars in a fixed de Sitter background is obtained. To calculate the N-point function in Schr\"{o}dinger field theory, the “in-in” formalism is  extended in the Friedmann-Lema\^itre-Robertson-Walker (FLRW) universe. We compute the three-point function for primordial scalar field fluctuation in the single field inflation by this in-in formalism. The results are the same as the three-point function in the Heisenberg picture.}

\begin{document}

\maketitle

%%%%%%%%%%%%%%%%%%%%%%%%%%%%%%%%%%%%%%%%%%%%%%%%%%
\section{Introduction}
%%%%%%%%%%%%%%%%%%%%%%%%%%%%%%%%%%%%%%%%%%%%%%%%%%

In the standard model of cosmology  structures are originated from the quantum fluctuation in the de Sitter inflationary background which become classic after the horizon crossing. The main information about these structures and the cosmic microwave background (CMB) comes from the statistical study of the data \cite{Ade:2015xua}. The power spectrum and bispectrum of the temperature in the CMB data and density fluctuation in the large scale structure are the two important quantities for statistical study. Consequently, on the theoretical side one has to derive these quantities to study and compare the models with observation. The two and three-point function of curvature perturbation are usually calculated in the infalationary models to study the two and three-point function of the temperature in the CMB.
\\

People usually use the Heisenberg picture in field theory of the curved space time to calculate the two or three-point function of the curvature perturbation. In this way, the in-in formalism of the Heisenberg field theory helps to evaluate the expectation values of products of fields at a fixed time \cite{Weinberg:2005vy}. In contrast to the S-matrix method conditions are not imposed on the fields at both very early and very late times, but only at very
early times, when the wavelength is deep inside the horizon. This was very useful to study non-Gaussianity of the inflationary models. \\

On the side of the field theory, there is Schr\"{o}dinger picture which the operators are time independent. This picture is the natural framework to study the entanglement between the fields \cite{Holman-entanglement, Martin-Martinez:2014gra}. The Schr\"{o}dinger representation is also   best suited to discussions of decoherence of cosmological perturbations and the quantum to classical transition in inflationary models \cite{Polarski:1995jg,Burgess:2014eoa}. To this end, many works have been done to study the entanglement effect and the quantum to classical transition in the inflationary models \cite{Martin:2015qta}. Nevertheless, the question arise as to whether one can develop in-in formalism in the Schr\"{o}dinger field theory and apply it to calculate the bispectrum of the curvature perturbation. Indeed, this in-in formalism helps to better study of the entanglement  and the quantum to classical transition effects on the non-Gaussianity of the CMB temperature. \\

The purpose of this paper is twofold. First we  introduce Schr\"{o}dinger picture in practical way in flat and then curved space-time. Second it will be shown how this picture helps us in calculation of spectrum and bi-spectrum inflationary background. To this end, we have to develop the in-in formalism for the Schr\"{o}dinger field theory. The structure of this paper is as follows. In section II we introduce the the Schr\"{o}dinger field theory in the flat space time and in section III it will be extended to the curved spacetime, and the scalar field power spectrum in de Sitter background will be calculated. Section IV is devoted to deriving the in-in formalism in Schr\"{o}dinger field theory which helps to calculate the expectation value of any physical quantity. Then, in section V we apply this in-in formalism for calculating the three-point function of a single field inflation. Finally, we conclude with a discussion in section VI.\\

%%%%%%%%%%%%%%%%%%%%%%%%%%%%%%%%%%%%%%%%%%%%%%%%%%
\section{Schr\"{o}dinger Picture In Flat Space-Time}
%%%%%%%%%%%%%%%%%%%%%%%%%%%%%%%%%%%%%%%%%%%%%%%%%%
In order to study behavior of a field in the FLRW universe which is a curved spacetime first, we introduce the Schr\"{o}dinger field theory in flat spacetime and then we extend it to the FLRW spacetime. Quantum field theory is nothing more than a quantum theory with a continuous set of degree of freedoms in which every field is impressed by some symmetries. To construct a quantum theory one needs the classical counterpart theory. Consider a dynamical system $\{q_1,q_2,\ldots ,q_n\}$ degrees of freedom. For  quantum dynamics they have the following quantization relations:
\begin{equation}
	[\hat{q_i},\hat{q_j}]=[\hat{p_i},\hat{p_j}]=0; \ \ \ \ \ \ [\hat{q_i},\hat{p_j}]=i \delta_{ij} ;
\end{equation} 
in which $\hat{p_j}$ is the canonical momentum  $\hat{q_j}$ . In the second step, we need to construct a Hilbert space. Vectors in such Hilbert space have to be square integrable respect to the inner product which the space is endowed by:
\begin{equation}
	\langle \Psi |\Psi \rangle<\infty; \ \ \ \ where\ \ \ \langle \Psi_1|\Psi_2 \rangle\equiv\int Dq \Psi_1(q_i)^* \Psi_2(q_i).
\end{equation}
For sake of convenience, we consider a limited distribution basis in the Hilbert space. However, this basis is not square integrable, so it would be very useful in calculations over the Hilbert space. Such basis are constructed as eigen states of $\hat{q_i}$ or $\hat{p_i}$ operators 
\begin{equation}
	\hat{q_i}|q\rangle=q_i|q\rangle;\ \ \ \ \ \ \hat{p_i}|p\rangle=p_i|p\rangle.
\end{equation} 
They are complete in the sense that
\begin{equation}
	\langle q|q^{\prime}\rangle=\delta(q-q^{\prime});\ \ \ \ \  \int Dq\  |q\rangle \langle q|=I.
\end{equation}
And the same for $|p\rangle$. It's so easy to investigate that:
\begin{equation}
	\langle q|p\rangle=\frac{1}{\sqrt{(2\pi)^n}}\exp(ip.q) \ ;\ \ \  \hat{p_i}=-i\frac{\partial}{\partial q_i}.
\end{equation}
Every physical state is just a vector in the Hilbert space which satisfies Schr\"{o}dinger equation 
\begin{equation}
	H(q_i, -i\frac{\partial}{\partial q_i}) \Psi(q_i)=i\frac{\partial\Psi}{\partial t}(q_i).
\end{equation}
It is straightforward to generalize above formalism to the case of any field by following rules:
\begin{equation}
	i\longrightarrow x; \ \ 
	\hat{q_i}\longrightarrow \hat{\phi}(x);\ \ 
	\hat{p_i}\longrightarrow \hat{\Pi}(x);\ \ 
	-i\frac{\partial}{\partial q_i}\longrightarrow -i\frac{\delta}{\delta \phi(x)};\ \ 
	\Psi(q_i)\longrightarrow \Psi[\phi].
\end{equation}
Here $x$ is no longer the dynamical variable and this responsibility is left to $\phi(x)$. For such quantum theory we need the same canonical relation as we had before
\begin{equation}
	[\phi(x),\phi(y)]=[\Pi(x),\Pi(y)]=0;\ \ \ [\phi(x),\Pi(y)]=i\delta(x-y).
\end{equation}
Note that we are in the Schr\"{o}dinger picture so all operators are time independent. Also as before, the Hilbert space may be square integrable in field configuration space:
\begin{equation}
	F[\phi] \in L^2\ \  if\ \ \ \int D\phi F[\phi]^* F[\phi] <\infty.
\end{equation} 
There are also canonical basis:
\begin{equation}
	\hat{\phi(x)}|\phi\rangle=\phi(x)|\phi\rangle;\ \ \langle \phi|\phi^{\prime}\rangle=\delta(\phi-\phi^{\prime});\ \ \ \ \  \int D\phi\  |\phi\rangle \langle \phi|=I.
\end{equation}
From mathematical point of view the integral measure i.e. $D\phi$ is not well defined but for practical calculation we will not encounter with any problem. We could also find the same relation as before 
\begin{equation}
	\langle \phi|\Pi\rangle=A\exp[i\int dx \phi(x)\Pi(x)] \ ;\ \ \  \hat{\Pi(x)}=-i\frac{\delta}{\delta \phi(x)}.
\end{equation}
Finally, we need the last chapter of this story i.e. Schr\"{o}dinger equation in the functional form
\begin{equation}
	H(\phi(x), -i\frac{\delta}{\delta \phi(x)}) \Psi(\phi)=\int dx \mathcal{H}(\phi(x), -i\frac{\delta}{\delta \phi(x)}) \Psi(\phi)=i\frac{\partial\Psi}{\partial t}(\phi).
\end{equation}
Note that, when the Hamiltonian is time independent, then the above equation would be reduced to 
\begin{equation}\label{indept}
	H(\phi(x), -i\frac{\delta}{\delta \phi(x)}) \Psi(\phi)=\int dx \mathcal{H}(\phi(x), -i\frac{\delta}{\delta \phi(x)}) \Psi(\phi)=E \Psi(\phi).
\end{equation}
In such case, time dependent part of wave function is just a phase factor $e^{iEt}$.
We made the Schr\"{o}dinger functional formalism in one dimension and without any difficulty, we could do this in any arbitrary dimension.
Usually, it's better to find a more convenient basis to do our calculations. If we impose that the fields live in a finite space with a volume $V$ (one dimensional case $L$) the following expansion of operators is reasonable
\begin{align}
	\hat{\phi}(x)=\frac{1}{\sqrt{L}}\sum_k \hat{\phi}_k e^{-ikx}\\
	\hat{\Pi}(x)=\frac{1}{\sqrt{L}}\sum_k \hat{\Pi}_k e^{-ikx},
\end{align}
where $k$ represents  the spacial momentum mode. If the field is real then we have:
\begin{align}
	\hat{\phi}_k^\dagger =\hat{\phi}_{-k},\\
	\hat{\Pi}_k^\dagger =\hat{\Pi}_{-k}.
\end{align}
Such transformation is invertible by using the Fourier integral. Hence, it's very simple to see that we have new commutation relation for new basis:
\begin{align} \label{com1}
	[\hat{\phi}_k,\hat{\phi}_{k^{\prime}}]=[\hat{\Pi}_k,\hat{\Pi}_{k^{\prime}}]=0; \\ \nonumber [\hat{\phi}_k,\hat{\Pi}_{k^{\prime}}]=i \delta(k+k^{\prime}). 
\end{align}
At the moment, we have all things to quantize a field for a given action. Let us write down the usual action for massive scalar field and examine above formalism:
\begin{align}
	S=\frac{1}{2}\int dt dx(\partial_{\mu}\phi \partial^{\mu}\phi -m^2\phi^2).
\end{align}
The Hamiltonian for this Lagrangian is 
\begin{align}
	H=\frac{1}{2}\int (\Pi^2+(\frac{d\phi}{dx})^2+m^2\phi^2) dx.
\end{align}
Using the new basis one could drive 
\begin{align}
	H=\frac{1}{2}\sum_k \Pi_k \Pi_{-k}+\omega^2_k\phi_k \phi_{-k},
\end{align}
in which $\omega_k^2=k^2+m^2$ is the energy for every degree of freedom. This Hamiltonian is nothing just as a combination of many decoupled harmonic oscillators in one dimension. We learned from quantum mechanics how this Hamiltonian is exact solvable. By using annihilation and creation operators for each mode we get
\begin{align}\label{a1}
	a_k=\frac{1}{\sqrt{2\omega_k}}(\Pi_k-i\omega_k\phi_{-k})
\end{align}
and
\begin{align}
	a_k^\dagger=\frac{1}{\sqrt{2\omega_k}}(\Pi_{-k}+i\omega_k\phi_{k}).
\end{align} 
We would like to rewrite the Hamiltonian in the more useful form
\begin{align}
	H=\sum_k \omega_k (a^\dagger_k a_k +\frac{1}{2}).
\end{align}
Using the commutation relations (\ref{com1}), for annihilation and creation operators we have 
\begin{align}\label{a2}
	[a_k,a_{k^{\prime}}]=[a_k^\dagger,a^\dagger_{k^{\prime}}]=0
\end{align}
or
\begin{align}\label{a4}
	[a_k,a^\dagger_{k^{\prime}}]=\delta(k+k^{\prime}).
\end{align}
This Hamiltonian acts on the Fock space. In practice, we find irreducible representation of the  Hamiltonian on the  Fock space. This Hamiltonian is positive definite and has a unique minimum. Practically, we just measure relative energy in experiments so we can drop out the  $\frac{1}{2}$ term in Hamiltonian and fix the minimum energy in zero value. We cannot neglect this term in the curved spacetime because this zero point or vacuum energy changes the spacetime geometry.
The corresponding state known as a vacuum $|0\rangle$ vector which
\begin{align}
	a_k |0\rangle=0,\\
	a_{k_n}^\dagger\ldots  a_{k_2}^\dagger a_{k_1}^\dagger |0>=|k_1,k_2,...,k_n\rangle.
\end{align}
Now, we are in a position which could find the time independent physical wave function $\Psi[\phi]$ in (\ref{indept}), in the modes space. Using (\ref{a1}) and (\ref{a2}) we get 
\begin{align} \label{gaussi}
	\Psi[\phi]=A\ \exp(-\frac{1}{2}\sum_k \omega_k\phi_k \phi_{-k}).
\end{align}
Note that the particle interpretation of such a quantization is encoded in $\phi$ not in $\Psi[\phi]$, and the wave function measures observable (Hermitian operators) like N-point functions. \\
One maybe thinks that the Gaussian solution (\ref{gaussi}) accidentally happens in the Minkowski space-time but it's not true. In the next section we will show this it the case for the all spatially maximal symmetric spacetime manifolds. It is important to mention that representation of Fock space as a  Schr\"{o}dinger  functional, depends on the choosing of the spacetime foliation. In the other words, if one constructs a  Schr\"{o}dinger  field theory in $\Sigma_1$ and then another one does on $\Sigma_2$ foliations, these two field theories are not unitary relevant \cite{quvedo2002}. Moreover, there is no  Schr\"{o}dinger  theory for non-globally hyperbolic spacetime. 

%%%%%%%%%%%%%%%%%%%%%%%%%%%%%%%%%%%%%%%%%%%%%%%%%%

%%%%%%%%%%%%%%%%%%%%%%%%%%%%%%%%%%%%%%%%%%%%%%%%%%
\section{Schr\"{o}dinger Picture In FLRW Space-Time}\label{c1}
In this section we shall quantize a free scalar field in a curved background. Here we consider de Sitter background and try write down a quantum field theory in Schr\"{o}dinger picture for this case. However one could find time-like killing vector in some patch, we prefer to do in a time dependent version of de Sitter background. The reason is that
the time dependent form of de Sitter background metric, has a cornerstone role in the first approximation of inflationary models. Let choose the chart such that the metric has the simple form
\begin{align}
	ds^2=-dt^2+a(t)^2d\vec{x}^2,\ \ \ \ a(t)=e^{Ht}
\end{align} 
where $H$ is constant Hubble parameter. We could write the conformal form of this metric 
\begin{align}
	ds^2=a(\eta)^2\left[-d\eta^2+d\vec{x}^2\right],\ \ \ \eta=-\frac{1}{H}e^{-Ht}\ \ \ \ for\ \ \ -\infty <\eta<0.
\end{align}
Now suppose a free scalar field in this background has the below action
\begin{align}\label{b1}
	S=\frac{1}{2}\int d^4x a^4(\eta)\left(\frac{1}{a^2(\eta)}[{\phi^{\prime}}^2(\eta,x)-(\nabla\phi(\eta,x))^2]-m^2\phi^2(\eta,x)\right).
\end{align}
Before going further, it would be useful to mention some notes about this action. Practically such an action appears in leading order of slow roll parameters, when one expands action of an scalar field $\Phi$ (the inflaton field) in (\ref{a20}), in term of curvature perturbation. In some papers like \cite{Maldacena:2002vr}, $\zeta$ is used as curvature perturbation which here in (\ref{b1}),  $\phi$ play the curvature perturbation role. 
The difference of (\ref{b1}) with first order term in \cite{Maldacena:2002vr} is just in re-scaling of expansion factor $a(\eta)$ which in our case should be multiply with $\frac{\dot{\phi}}{H}$ to be the same of \cite{Maldacena:2002vr}. Note that there, $\phi$ has the role of inflaton field.
We need the Hamiltonian in order to set up our dynamical system in the Schr\"{o}dinger picture. The Hamiltonian is 
\begin{align}
	H_{\phi}=\int d^3x \{\frac{\pi^2_{\phi}}{2a^2}+\frac{a^2}{2}({\nabla\phi}^2+a^2m^2_{\phi}\phi^2) \}.
\end{align}
It will be more useful to write down the Hamiltonian in terms of the comoving spatial momentum modes. Therefore, we can decompose $\phi$ and its conjugate momentum as follows
\begin{align}
	\phi(x)=\frac{1}{\sqrt{V}}\sum_k \phi_k e^{-ik.x}\\
	\pi(x)=\frac{1}{\sqrt{V}}\sum_k \pi_{k,\phi} e^{-ik.x}.
\end{align}
In terms of these modes we have
\begin{align}
	H_{\phi}=\frac{1}{2}\sum_k (\frac{\pi_{k,\phi} \pi_{-k,\phi}}{a^2}+a^2(k^2+m_{\phi}^2a^2)\phi_k \phi_{-k}).
\end{align}
It's just like a harmonic oscillator with time dependent coefficients. Therefore, it will be natural to propose a solution with unknown coefficients and try to examine it such that it satisfies Schr\"{o}dinger equation. Note that, because the Hamiltonian is separable for each mode, we could write the total solution as 
\begin{align}
	\psi_k[\phi_k;\eta]\equiv\langle \phi_k |\Psi[\phi;\eta]\rangle \\
	\Psi[\phi;\eta]=\Pi_k \psi_k[\phi_k;\eta]
\end{align}
which $\psi_k$ is the wave function for corresponding $k$-Hamiltonian mode such that satisfies Schr\"{o}dinger equation 
\begin{align}
	i\partial_{\eta}\psi_k[\phi_k;\eta]=H_k\psi_k[\psi_k;\eta].
\end{align}
Because of quadratic form of Hamiltonian, our ansatz for solution will be Gaussian
\begin{align}
	\psi_k[\phi_k;\eta]=N_{k}(\eta)exp[-\frac{1}{2}A_k(\eta)\phi_k\phi_{-k}].
\end{align}
Inserting the ansatz into the Schr\"{o}dinger equation and then matching the power field modes in RHS and LHS gives us two differential equations as follows 
\begin{align}
	i\frac{N_k^{\prime}}{N_k}&=\frac{A_k}{2a^2}\\
	iA^{\prime}_k&=\frac{A_k^2}{a^2}-\omega_{k}(\eta)^2a^2,\ \ \ \ \ \omega_k(\eta)^2=k^2+m^2 a^2(\eta).
\end{align}
The equation for $A_k$ is just a Riccatti equation and one could transform it to a second order (ordinary) differential equation by writing 
\begin{align}\label{Ak}
	iA_k(\eta)=a^2(\eta)\left(\frac{f^{\prime}_k(\eta)}{f_k(\eta)}-\frac{a^{\prime}(\eta)}{a(\eta)}\right).
\end{align}
The resulting equation will be 
\begin{align}\label{a7}
	f^{\prime\prime}_k+\left(\omega_{k}^2(\eta)-\frac{a^{\prime\prime}}{a}\right)f_k=0.
\end{align}
It's so simple to show that the above equation is nothing but a Bessel function solution. To see this, let write (\ref{a7}) in the exact form of conformal time 
\begin{align}\label{a8}
	f^{\prime\prime}_k+\left(k^2-(2-\frac{m^2}{H^2})\frac{1}{\eta^2}\right)f_k=0.
\end{align}
Now if we use new variable $f_k=\sqrt{|\eta|k}v_k$, then the resulting will be as follows
\begin{align}
	s^2v_k^{\prime \prime}+s v_k^{\prime}+\left(s^2-n^2\right)v_k=0,\ \ \ s=k\eta;\ \ \ n=\sqrt{\frac{9}{4}-\frac{m^2}{H^2}}.
\end{align}
Here we assume that $m\ll H$.
The general solution (\ref{a8}) is 
\begin{align} \label{hanckel}
	f_k(\eta)=\sqrt{k|\eta|}\left(\alpha J_n(k|\eta|)+\beta Y_n(k|\eta|)\right).
\end{align}
One may use Wronskian $W(f_k,f_k^*)=-i$ (which comes from the extra degree of freedom in Riccatti transformation of initial condition) and finds a constraint 
\begin{align}
	\alpha\beta^*-\beta\alpha^*=\frac{i\pi}{k}.
\end{align}
But it's not sufficient to find $\alpha$ and $\beta$. We will need some asymptotic conditions to determine these coefficients uniquely. These asymptotic conditions known as Bunch-Davies condition 
\begin{align}
	f_k(\eta)\rightarrow \frac{1}{\sqrt{\omega_k}}e^{i\omega_k},\ \ \frac{f_k^{\prime}}{f_k}\rightarrow i\omega_k \ \ \ as\ \ \eta\rightarrow -\infty.
\end{align}
With assumption of these conditions we find 
\begin{align}
	f_k^{BD}(\eta)=\frac{\sqrt{\pi|\eta|}}{2}\left(J_n(k|\eta|)-iY_n(k|\eta|)\right)=\frac{\sqrt{\pi|\eta|}}{2} H_n^{(1)}(k|\eta|).
\end{align}
At the moment, we can calculate the $\psi_k$. Without any problem we are able to find density matrix which has all information about what we want to calculate, namely N-point functions. The resulting density matrix is 
\begin{equation}
	\rho=|\Psi[\phi;\eta]\rangle \langle \Psi[\phi;\eta] |.
\end{equation}
Since our wave function lives on an infinite dimensional manifold with coordinate which labeled by $\phi$, components of such a density matrix is in the following form
\begin{equation}
	\rho[\phi,\tilde{\phi};\eta]=\Pi_{k}\langle \phi_k |\Psi[\phi;\eta]\rangle \langle \Psi[\phi;\eta] |\tilde{\phi}_k\rangle.
\end{equation}
For the reason that each positive mode there exists a negative one  (which has the same wave function and also dependency of each mode), one could  write down the density matrix as a product of all positive modes. Each positive mode has the following form of density matrix
\begin{equation}
	\rho_k[\phi_k, \tilde{\phi_k};\eta]=|N_{k}(\eta)|^4 exp[-A_k(\eta)\phi_k\phi_{-k}-A^{*}_k(\eta)\tilde{\phi}_k\tilde{\phi}_{-k}].
\end{equation}
We could normalize this density matrix such that its trace equals to one 
\begin{equation}
	\int \mathcal{D}\phi_k \mathcal{D}\phi_{-k}\rho[\phi_k,\phi_k;\eta]=|N_k|^4 \int  exp[-2A_{Rk}(\eta)\phi_k\phi_{-k}]=\frac{2\pi}{2 A_{Rk}}|N_k|^4=1,
\end{equation}
where the $ A_{Rk} $ is the real part of $ A_{k} $. Using equation (\ref{Ak}) and the Wronskian equation we get
\be
A_{Rk}=\frac{a^2}{2|f_k|^2}.
\ee\\

Now, we are in a position to calculate any N-point function. Because of Gaussian form of density matrix, it is easy to argue that all of odd-number correlation functions are equal to zero. Hence it's enough to calculate even-number correlation functions. For our purpose, we just need to calculate two-point correlation function which appears in quantum fluctuations in CMB. One could see in this calculation $<\phi_{k_1}\phi_{k_2}>$ has a non-vanishing value when $k_1=-k_2$ and this value is as follows
\begin{equation}
	<\phi_{k_1} \phi_{k_{2}}>=\Pi_{k} |N_k|^4\int \mathcal{D}^2 \phi_k \left(\phi_{k_1} \phi_{k_{2}}\right) exp[-2A_{Rk}(\eta)\phi_k\phi_{-k}]=\frac{1}{2A_{Rk_1}}\delta (k_1+k_2),
\end{equation}
where $\mathcal{D}^2 \phi_k= \mathcal{D}\phi_k \mathcal{D}\phi_{-k}$. The power spectrum 
\be
\Delta^2_\phi(k)=\frac{k^3}{2\pi^2}<\phi_k \phi_{-k}>|_{\eta\rightarrow 0}.
\ee
As shown in \cite{Holman-entanglement} in the case of massless scalar field, $f_k$ in (\ref{hanckel}), is just the Hanckel function of order $3/2$. This would have power-law behavior, $\Delta^2_\phi(k)\propto k^{(n_s-1)}$, which in the free field case the spectral index is $n_s=1$, which shows the scale independency of the de Sitter background.\\

Here recall that the application of Schr\"{o}dinger field theory in inflationary universe is different from the flat spacetime which is used in particle physics. The field theory in flat spacetime uses in-out initial condition (past and late time) to calculate the physical quantities like N-point function for S-matrix. But in cosmology we have the past time initial condition and the physical quantity are calculated in a finite time (like horizon crossing time for a wavelength). As a result, it is needed to develop another formalism to calculate the  physical quantities like N-point function other than S-matrix which is called in-in formalism. In the following section, we describe how to apply the interaction Hamiltonian and the density matrix of the free part to calculate the expectation value of a physical quantity in the in-in formalism of the Schr\"{o}dinger field theory.\\

%%%%%%%%%%%%%%%%%%%%%%%%%%%%%%%%%%%%%%%%%%%%%%%%%%

%%%%%%%%%%%%%%%%%%%%%%%%%%%%%%%%%%%%%%%%%%%%%%%%%%
\section{in-in Formalism}
Up to now, we've learned how to investigate the wave function of a quantum theory in a curved space-time. This wave function is a functional of fields in Fourier basis which we have chosen previously. However, we found the wave function in the case of free scalar field, it is no difficulty to calculate N-point function when system has an interaction. Here, we find a method for this kind of calculations, nevertheless the method needs perturbation approach to calculate the expectation value of the physical quantities. We call this method in-in formalism here is useful to mention another approach to this problem which gives the same result (see Appendix B for short review). Now, suppose $H_0[\phi;t]$ is a quadratic Hamiltonian (in $\phi$) and $H_i [\phi;t]$ be the interaction part which one wants to add to the background (which of course, includes all cubic and higher powers of $\phi$ ). Starting from Schr\"{o}dinger equation
\begin{equation}
	i\dot{\Psi}(\phi;t)=H[\phi;t]\Psi(\phi;t)=\left[H_0+H_i\right]\Psi(\phi;t)
\end{equation}
and then pass through the interaction picture, one can define a wave function in such picture as
\begin{equation}
	\Psi(t)^{\mathcal{I}}\equiv U_0^{-1}(t,t_0)\Psi(t)
\end{equation}
which we defined $U_0(t,t_0)$ as 
\begin{equation}\label{a13}
	\frac{d}{dt}U_0(t,t_0)=-i\,H_0\,U_0(t,t_0).
\end{equation}
Then by time derivation of $\Psi^{\mathcal{I}}$, we find
\begin{equation}
	i\dot{\Psi}^{\mathcal{I}}(t)=H_{I}\Psi^{\mathcal{I}}(t),
\end{equation}
where $H_I$ is defined as follows
\begin{equation}
	H_I \equiv U_0(t,t_0)H_iU_0^{-1}(t,t_0).
\end{equation}
Using recursive integrating we find 
\begin{equation}
	\Psi^{\mathcal{I}}(t)= T exp\left[-i\int H_I(t^{'})dt^{'}\right]\Psi^{\mathcal{I}}(t_0)=F(t,t_0)\, \Psi^{\mathcal{I}}(t_0)
\end{equation}
and then one could find the density matrix as follows
\begin{equation}
	\rho(t)=\Psi(t)\Psi^{\dagger}(t)=U_0 (t,t_0)\, F(t,t_0)\, \rho(t_0)\, F^{-1}(t,t_0)\,  U_0^{-1} (t,t_0) .
\end{equation}
Now, we have a formula for density matrix in any time then one could  find expectation value of any functional of $\phi_{k}$'s and $\pi_{k}$'s as following
\begin{equation}
	\langle \mathcal{K}[\phi_k, \pi_k]\rangle =Tr\left[\mathcal{K}[\phi_k, \pi_k]\rho(t)\right] = Tr\left[F^{-1} (t,t_0)\, \mathcal{K}_{\mathcal{I}}[\phi_k, \pi_k;t,t_0]\, F(t,t_0)\, \rho(t_0)\right]\ ,
\end{equation}
where
\begin{equation} \label{int}
	\mathcal{K}_{\mathcal{I}}[\phi_k, \pi_k;t,t_0]=U_0^{-1} (t,t_0)\, \mathcal{K}[\phi_k, \pi_k]\, U_0 (t,t_0)
\end{equation}
Using this relation, we can calculate the expectation value of every physical operator like power spectrum and bispectrum, by replacing in $\mathcal{K}$. Such values could not be evaluated exactly and we have to expand $F(t,t_0)$ in term of $H_I$. By expanding this relation in the first order of $H_I$ one finds 
\begin{equation}
	\langle \mathcal{K}[\phi_k , \pi_k] \rangle = -i \int_{t_0}^{t} dt^{'} Tr ([ \mathcal{K}_{\mathcal{I}}[\phi_k, \pi_k;t,t^{'}],H_{i}(t_0)] \rho_0 (t')).
\end{equation}

Note that in this relation $\rho_0 (t^{'})$ is the density matrix for quadratic Hamiltonian $H_0$, which has been defined as 
\begin{equation}
	\rho_0 (t') = U_0 (t^{'},t_0)\, \rho(t_0)\, U_0^{-1} (t^{'},t_0).
\end{equation}\\

%%%%%%%%%%%%%%%%%%%%%%%%%%%%%%%%%%%%%%%%%%%%%%%%%%%%%%%%%%%%%
\section{Application for the inflation}
%%%%%%%%%%%%%%%%%%%%%%%%%%%%%%%%%%%%%%%%%%%%%%%%%%%%%%%%%%%%%

Now, we are in a position which allows us to work out such in-in formalism for a given action in the Schr\"{o}dinger field theory. In this section, calculation will be done in the case of inflationary universe which the cubic terms of action should  be considered. These cubic terms would affect on three-point functions which are called non-Gaussianities. These non-Gaussianties could be important when we have two scalar fields in the early universe.
In such Hamiltonian one could see some parameters. To explain what they are, we have to explain the action which leads to (\ref{ham-in}) 
\begin{equation} \label{a20}
	S=\frac{1}{2}\int d^4x \sqrt{-g} \left[M_{pl}^2R +
	2P(X,\Phi)\right]~,
\end{equation}
where $P(X,\Phi)$ involves Lagrangian of a scalar field in an arbitrary potential and here $X=-\frac{1}{2} g_{\mu \nu}\partial_{\mu}\Phi \partial_{\nu}\Phi$. Then one could  defined $c_s$ (like speed of sound) as  

\begin{eqnarray}
	c_s^2 = \frac{dP}{dE}= \frac{P_{,X}}{P_{,X}+2X P_{,XX}}
\end{eqnarray}
where 

\begin{equation}
	E=2X P_{,X} -P ~.
\end{equation}

Finally, the rest of parameters are defined as follows
\begin{equation}
	\Sigma\equiv \frac{H^2\epsilon}{c_s^2}
\end{equation}

\begin{equation}
	s=\frac{\dot{c}_s}{c_s H}
\end{equation}

\begin{equation}
	\lambda\equiv X^2 P_{,XX}+\frac{2}{3}X^3 P_{,XXX}.
\end{equation}

Here, we use the Hamiltonian of the cubic terms as \cite{Chen:2009zp}. This Hamiltonian is as follows  
\begin{eqnarray} \label{ham-in} 
	H_{int}(t)&=&-\int
	d^3x\{ -a^3 (\Sigma(1-\frac{1}{c_s^2})+2\lambda)\frac{\dot{\phi}^3}{H^3}+
	\frac{a^3\epsilon}{c_s^4}(\epsilon-3+3c_s^2)\phi\dot{\phi}^2
	\nonumber \\ &+&
	\frac{a\epsilon}{c_s^2}(\epsilon-2s+1-c_s^2)\phi(\partial\phi)^2-2a\frac{\epsilon}{c_s^2}\dot{\phi}(\partial
	\phi)(\partial \chi) \}~
\end{eqnarray}
where $\epsilon$ is the slow roll parameter and $\chi$ is related non-locally to momentum 
\begin{equation}
	\partial^2\chi=a^2\frac{\epsilon}{c_s^2}\dot{\phi}.
\end{equation}
\footnote{Note that, there is an extra term in the cubic Hamiltonian which comes from  redefinition of the curvature perturbation in the expansion of the action. Here, we neglect this term in our typical calculations, however one can show that the contribution of that term in the Schr\"{o}dinger picture is equal that of found in \cite{Chen:2009zp}}. The physical interpretation of these parameters are not important here and one could find t in (\cite{Chen:2009zp}).
In leading order of perturbation, the three-point  correlation function can be calculated in following way

\begin{equation}
	\langle  \phi_{k_1} \phi_{k_2} \phi_{k_3} \rangle = -i \int_{t_0}^{t} dt^{'} Tr ([ {\phi_{k_1}}_{\mathcal{I}} {\phi_{k_2}}_{\mathcal{I}} {\phi_{k_3}}_{\mathcal{I}},H_{int}(t_0)] \rho(t^{'})).
\end{equation}
Here, ${\phi_k}_{\mathcal{I}}$ is defined as (\ref{int}). There are some points which should to be mentioned. First note that to do such calculation, one needs to work out the Fourier transformation of the Hamiltonian. Here, we have divided the Hamiltonian into four parts and then find the Fourier transformation of each part, $ H_{int}=H_{int}^{(1)}+ H_{int}^{(2)}+ H_{int}^{(3)}+ H_{int}^{(4)} $.
For the first part we have
\begin{eqnarray}
	H_{int}^{(1)}&=&-\int
	d^3x\{ -a^3 (\Sigma(1-\frac{1}{c_s^2})+2\lambda)\frac{\dot{\phi}^3}{H^3} \nonumber \\ &=& 	 \frac{1}{(2\pi)^{3/2}}\Sigma_{k,k^{'},k^{''}} \frac{a^{-6}}{H^3} \{\Sigma(1-\frac{1}{c_s^2})+2\lambda\}\pi_{k}\pi_{k^{'}}\pi_{k^{''}}\delta^{3}(k+k^{'}+k^{''})
\end{eqnarray}
which delta function arises from integration over whole space.
For the second part we have
\begin{eqnarray}
	H_{int}^{(2)}&=&  -\int d^3x \frac{a^3\epsilon}{c_s^4}(\epsilon-3+3c_s^2)\phi\dot{\phi}^2 \nonumber \\ &=& -\frac{1}{(2\pi)^{3/2}}\Sigma_{k,k^{'},k^{''}}\frac{a^{-3}\epsilon}{c_s^4}(\epsilon-3+3c_s^2)\phi_{k}\pi_{k^{'}}\pi_{k{''}}\delta^{3}(k+k^{'}+k^{''}) ,
\end{eqnarray}
for the third part
\begin{eqnarray}
	H_{int}^{(3)}&=&-\int d^3x \frac{a\epsilon}{c_s^2}(\epsilon-2s+1-c_s^2)\phi(\partial\phi)^2 \nonumber \\ &=& -\frac{1}{(2\pi)^{3/2}}\Sigma_{k,k^{'},k^{''}} \frac{a\epsilon}{c_s^2}(\epsilon-2s+1-c_s^2)\vec{k^{''}}.\vec{k^{'}} \phi_k \phi_{k^{'}} \phi_{k^{''}}\delta^{3}(k+k^{'}+k^{''})
\end{eqnarray}
and finally for the last part one can find
\begin{equation}
	H_{int}^{(4)}=\int d^3 x 2a\frac{\epsilon}{c_s^2}\dot{\phi}(\partial
	\phi)(\partial \chi)=\frac{1}{(2\pi)^{3/2}}\Sigma_{k,k^{'},k^{''}} \frac{2\epsilon^2}{c_{s}^4 a^3}\frac{\vec{k^{'}}. \vec{k^{''}}}{(k^{''})^2}\phi_k \pi_{k^{'}}\pi_{k^{''}}\delta^{3}(k+k^{'}+k^{''}).
\end{equation}
The second point is that, for calculating of such N-point functions one encounters some integration over configuration of field space like we have seen in path integral method. We have worked out two-point one in section \ref{c1}, and here we just need to do it for three-point one. It is not hard to show that 
\begin{eqnarray}
	\prod_k \int \mathcal{D}^2 \phi_{k} \left(\phi_{k_{1}} \phi_{k_2} \phi_{k_3} \phi_{k_4}\right) exp[-2A_{Rk}(\eta)\phi_k\phi_{-k}] \thicksim \nonumber \\ \delta(k_1+k_2)\delta(k_4+k_3)\frac{1}{2A_{Rk_1}}\frac{1}{2ARk_3}+2\leftrightarrow 3 +2 \leftrightarrow 4.
\end{eqnarray}
To reach an exact solution, we fix time integration  from $\eta=-\infty$, the birth of inflation, to $\eta=0$ the end of inflation. In addition, we are going to suppose the field in action (\ref{b1}) is a massless one. Then the first part of Hamiltonian  yields 
\begin{eqnarray}
	\langle  \phi_{k_1} \phi_{k_2} \phi_{k_3} \rangle &=& -i \int_{-\infty}^{0} d\eta\  a(\eta) Tr ([{\phi_{k_1}}_{\mathcal{I}} {\phi_{k_2}}_{\mathcal{I}} {\phi_{k_3}}_{\mathcal{I}},H^{(1)}_{int}] \rho)\nonumber \\ &=&\alpha\  \delta^{3} (k_1+k_2+k_3) \frac{(k_1 k_2 k_3)^2}{(k_1+k_2+k_3)^3} \frac{1}{(k_1 k_2 k_3)^3}
\end{eqnarray}

where

\begin{equation}
	\alpha=-(2\pi)^3 \frac{3H^4}{8\epsilon^2 c_s^4}(c_s^2-1+\frac{2\lambda c_s^2}{\Sigma}).
\end{equation}
For the second part one finds

\begin{eqnarray}
	\langle  \phi_{k_1} \phi_{k_2} \phi_{k_3} \rangle =\langle  \phi_{k_1} \phi_{k_2} \phi_{k_3} \rangle = -i \int_{-\infty}^{0} d\eta\  a(\eta) Tr ([{\phi_{k_1}}_{\mathcal{I}} {\phi_{k_2}}_{\mathcal{I}} {\phi_{k_3}}_{\mathcal{I}},H^{(2)}_{int}] \rho)=\\ \nonumber 
	\beta\  \delta^{3} (k_1+k_2+k_3) \frac{1}{(k_1 k_2 k_3)^3} Per\left(\frac{(k_2 k_3)^2}{(k_1+k_2+k_3)}+\frac{(k_1 k_2 k_3)^2}{k_1 (k_1+k_2+k_3)^2}\right)
\end{eqnarray}
where
\begin{equation}
	\beta=(2\pi)^3 \frac{H^4}{16\epsilon^2 c_s^4}(\epsilon-3+3c_s^2)
\end{equation}
where Per  means permutation of index 1 to 3. For the third part we have
\begin{eqnarray}
	& \langle  \phi_{k_1} \phi_{k_2} \phi_{k_3} \rangle = 
	\langle  \phi_{k_1} \phi_{k_2} \phi_{k_3} \rangle = -i \int_{-\infty}^{0} d\eta\  a(\eta) Tr ([{\phi_{k_1}}_{\mathcal{I}} {\phi_{k_2}}_{\mathcal{I}} {\phi_{k_3}}_{\mathcal{I}},H^{(3)}_{int}] \rho)= \\ \nonumber 
	& \gamma\  \delta^{3} (k_1+k_2+k_3) \frac{1}{(k_1 k_2 k_3)^3} Per\left( (k_1 . k_2)(-k_1-k_2-k_3 +\frac{k_1 k_2 +k_1 k_3 k_2 k_3}{k_1+k_2+k_3}+\frac{k_1 k_2 k_3}{(k_1+k_2+k_3)^2})\right)
\end{eqnarray}
which
\begin{equation}
	\gamma=(2\pi)^3 \frac{H^4}{16\epsilon^2 c_s^4}(\epsilon-2s+1-c_s^2)
\end{equation}
and finally for the last part, we find
\begin{eqnarray}
	&\langle  \phi_{k_1} \phi_{k_2} \phi_{k_3} \rangle = \langle  \phi_{k_1} \phi_{k_2} \phi_{k_3} \rangle = -i \int_{-\infty}^{0} d\eta\  a(\eta) Tr ([{\phi_{k_1}}_{\mathcal{I}} {\phi_{k_2}}_{\mathcal{I}} {\phi_{k_3}}_{\mathcal{I}},H^{(4)}_{int}] \rho)=\\ \nonumber 
	&-(2\pi)^3 \frac{H^4}{16\epsilon^2 c_s^4} \delta^{3} (k_1+k_2+k_3) \frac{1}{(k_1 k_2 k_3)^3} Per \left( \frac{k_1 . k_2}{(k_1+k_2+k_3)}(2k_3^2 +\frac{k_3^2 k_1+ k_3^2 k_2}{(k_1+k_2+k_3)})\right).
\end{eqnarray}
In the above results, we also use the commutation relation $[\phi_k,\pi_{k^{'}}]=\delta^3 (k+k^{'})$.
These results are in agreement with the results which found by others in \cite{Chen:2009zp, Maldacena:2002vr}.\\

%%%%%%%%%%%%%%%%%%%%%%%%%%%%%%%%%%%%%%%%%%%%%%%%%%
\section{Conclusion}
%%%%%%%%%%%%%%%%%%%%%%%%%%%%%%%%%%%%%%%%%%%%%%%%%%

In this work, after reviewing the Schr\"{o}dinger field theory in curved spacetime, we have tried to apply it for the inflationary models. First, we have used this formalism to calculate the power spectrum of non-interacting, massive minimally coupled scalars in a fixed de Sitter background. Second, we study the N-point function calculation which is an interesting quantity in the cosmological data, for instance, three-point function which is used to quantify the non-Gaussianity in the cosmology.\\

Since the initial condition in the inflation is given only at the early time with Bounch-Davies vacuum, we have to develop a formalism such as in-in formalism other than S-matrix which does not need late time condition. In this way we develop the in-in formalism for Schr\"{o}dinger field theory in curved space time. At the next step, we have calculated the three-point function for single field inflationary model. We have shown that this method gives the same results of in-in formalism of the Heisenberg field theory.\\

The main advantage of this formalism is that we can have a natural framework to study the entangled state effect on cosmological observation. One can also investigate the quantum to classical transition and decoherence of the state in the formalism to probe the inflationary models prediction of the cosmological date.\\

%%%%%%%%%%%%%%%%%%%%%%%%%%%%%%%%%%%%%%%%%%%%%%%%%%%%%%%%
{\bf Acknowledgments:}
\\

We would like to thank M.M. Sheikh-Jabbari and Hassan Firouzjahi for useful comments.\\

%%%%%%%%%%%%%%%%%%%%%%%%%%%%%%%%%%%%%%%%%%%%%%%%%%%%%%%%
\appendix
%%%%%%%%%%%%%%%%%%%%%%%%%%%%%%%%%%%%%%%%%%%%%%%%%%%%%%%%

\section{Gaussian Integrals}
In this appendix we're going to recall calculation method which we have used in computation of N-point functions based on Schr\"{o}dinger picture. We've seem that in scalar field theory case, the wave function is a functional of fields in Fourier basis. This wave function has a Gaussian from, so there is no difficulty to use Gaussian integrals which could be rewritten in functional forms. Starting from the one dimensional Gaussian integral as follows
\begin{equation}
	\int_{-\infty}^{+\infty}dx e^{-\frac{1}{2}a x^2}=(2\pi)^{\frac{1}{2}}a^{-\frac{1}{2}}
\end{equation} 
for positive real number $a$, one  can generalize this result to the N-dimensional one. Suppose $A$ is a symmetric and positive matrix, then by choice of a suitable basis one would be able to diagonalize $A$ and find the following result 
\begin{equation}
	\int dx_1dx_2\ldots dx_n exp\left[-\frac{1}{2}Y^{t}AY\right]=(2\pi)^{n/2}Det(A)^{-\frac{1}{2}}=(2\pi)^{n/2}e^{-\frac{1}2{}Tr Ln(A)}.
\end{equation}
We might generalize this case to the one which the number of variables is continuous. By absorbing the normalization factor in integral measure, it can be seen that
\begin{equation}
	\int \mathcal{D}\phi\  exp\left[\int dx dx^{'}\phi(x^{'})A(x,x^{'})\phi(x)\right]=Exp\left[{-\frac{1}2{}Tr Ln(A)}\right],
\end{equation} 
where $\phi(x)$ is a real scalar field. To compute any N-point function we just need even n-differentiation of the below integral
\begin{equation}
	\int dx_1dx_2\ldots dx_n exp\left[-\frac{1}{2}Y^{t}AY+\rho^t Y\right]=e^{\frac{1}{2}\rho^t A^{-1}\rho}\ e^{-\frac{1}2{}Tr Ln(A)}
\end{equation}
respect to $\rho$ arrays and then equal it  to zero value. Therefore, we have
\begin{eqnarray} \label{x1}
	\int dx_1dx_2\ldots dx_n(x_{i_1}x_{i_2}\ldots x_{i_n}) exp\left[-\frac{1}{2}Y^{t}AY\right]= \nonumber \\ e^{-\frac{1}2{}Tr Ln(A)}\left[A_{i_1 i_2}^{-1}A_{i_3 i_4}^{-1}\cdots A_{i_{n-1} i_{n}}^{-1}+Permutations\right]
\end{eqnarray}
The counterpart of this formula in the case of continuous dimension is as follows 
\begin{eqnarray}\label{p1}
	\int \mathcal{D}\phi \left(\phi(x_{i_1})\phi(x_{i_2})\ldots \phi(x_{i_n})\right) exp\left[\int dx dx^{'}\phi(x^{'})A(x,x^{'})\phi(x)\right]=\\ \nonumber e^{-\frac{1}2{}Tr Ln(A)}\left[A^{-1}(x_{i_1},x_{ i_2})A^{-1}(x_{i_3},x_{ i_4})\cdots A^{-1}(x_{i_{n-1}} ,x_{i_{n}})+Permutations\right].
\end{eqnarray}
The result could be generalize in a straight forward way for the case of complex scalar field. Just remember that for a discrete number of variable we have
\begin{equation} \label{z1}
	\int dz_1dz_1^* dz_2^*\ldots dz^*_n exp\left[-\frac{1}{2}Z^{\dagger}A Z\right]=(2\pi)^{n}Det(A)^{-1}=(2\pi)^{n}e^{-Tr Ln(A)}.
\end{equation}
To calculation of any N-point function in this case, one can differentiate respect to elements of $A$ and re-derive a formula like (\ref{x1}). For example, in case of two-point function we have
\begin{equation}
	\int dz dz^{*} z z^{*} exp(-\frac{1}{2}a z z^{*})=4 (2\pi) a^{-2}.
\end{equation}
Note that, the expectation value of $z^n$ is zero. To have non-zero value we need pair of $z$ and  $z^{*}$ in integral.
The functional counterpart of (\ref{z1})  is 
\begin{equation}
	\int \mathcal{D}\phi\ \mathcal{D}\phi^*\  exp\left[\int dx dx^{'}\phi^*(x^{'})A(x,x^{'})\phi(x)\right]=Exp\left[{-Tr Ln(A)}\right].
\end{equation}
and every N-point function is like (\ref{p1}) but without $1/2$ in exponent. Note that there  must be even number of both $\phi$ and $\phi^*$ to have  non-zero answers.

\section{Some Notes About in-in Formalism}
As we saw in section IV, to calculate the expectation value of any physical quantity, we need to have the density matrix operator in any time. In the Heisenberg picture, the in-in \cite{Weinberg:2005vy} formalism helps to calculate the  time evolution of any given operation like density matrix. Here, we review the in-in formalism in Heisenberg picture which gives the same results as we derived in the Schr\"{o}dinger  picture.\\
Let us recall perturbation method of the Hamiltonian formalism. Suppose we have some fields $\phi_a(x,t)$ and their conjugate momentum of them $\pi_a(x,t)$ which label $a$ just shows which kind of field here is assumed.
From canonical quantization point of view, we have usual commutation relations 
\begin{equation} 
	\Big[\phi_a({\bf x},t),\pi_b({\bf y},t)\Big]=i\delta_{ab}\delta^{3}({\bf x}-{\bf y})\;,~~~~~\Big[\phi_a({\bf x},t),\phi_b({\bf y},t)\Big]= \Big[\pi_a({\bf x},t),\pi_b({\bf y},t)\Big]=0\;,
\end{equation} 
and the equations of motion
\begin{equation} 
	\dot{\phi}_a({\bf x},t)=i\Big[H[\phi(t),\pi(t)],\phi_a({\bf x},t)\Big] \;,~~~~~~~ \dot{\pi}_a({\bf x},t)=i\Big[H[\phi(t),\pi(t)],\pi_a({\bf x},t)\Big] \;.
\end{equation} 
On the other hand, we know if there are some classical fields (which may arise from expectation value of quantum fields) then their dynamics obey from classical dynamics as follows 
\begin{equation} 
	\dot{\bar{\phi}}_a({\bf x},t)=\frac{\delta H[\bar{\phi}(t),\bar{\pi}(t)]}{\delta \bar{\pi}_a({\bf x},t)}\;,~~~~~
	\dot{\bar{\pi}}_a({\bf x},t)=-\frac{\delta H(\bar{\phi}(t),\bar{\pi}(t)]}{\delta \bar{\phi}_a({\bf x},t)}\;,
\end{equation} 
which here $\bar{\pi}_a({\bf x},t)$ and $\bar{\phi}_a({\bf x},t)$ are classical fields. 
Now, for sake of convenience, we assume quantum fluctuations around these classical quantities
\begin{equation} 
	\phi_a({\bf x},t)=\bar{\phi}_a({\bf x},t)+\delta\phi_a({\bf x},t)\;,~~~~~~~~~~
	\pi_a({\bf x},t)=\bar{\pi}_a({\bf x},t)+\delta\pi_a({\bf x},t)\;.
\end{equation}
We should keep in the mind that $\bar{\pi}_a({\bf x},t)$ and $\bar{\phi}_a({\bf x},t)$ are c-numbers and of course commute with anything. Hence, we have the same canonical commutation relations for perturbations 
\begin{equation} 
	\Big[\delta\phi_a({\bf x},t),\delta\pi_b({\bf y},t)\Big]=i\delta_{ab}\delta^{3}({\bf x}-{\bf y})\;,~~~~~\Big[\delta\phi_a({\bf x},t),\delta\phi_b({\bf x},t)\Big]= \Big[\delta\pi_a({\bf x},t),\delta\pi_b({\bf x},t)\Big]=0\;.
\end{equation}
One can expand the  Hamiltonian around such classical fields as follows
\begin{eqnarray} 
	H[\phi(t),\pi(t)]&=&H[\bar{\phi}(t),\bar{\pi}(t)]+ \sum_a\frac{\delta H[\bar{\phi}(t),\bar{\pi}(t)]}{\delta \bar{\phi}_a({\bf x},t)}\delta\phi_a({\bf x},t]
	+\sum_a\frac{\delta H[\bar{\phi}(t),\bar{\pi}(t)]}{\partial \bar{\pi}_a({\bf x},t)}\delta\pi_a({\bf x},t)
	\nonumber\\&&~~~~~~+\tilde{H}[\delta\phi(t),\delta\pi(t);t]\;,
\end{eqnarray} 
which $\tilde{H}$ is compact form of expansion terms in second and higher orders of perturbations. Although $H$ is the generator of time translation for $\phi_a$ and $\pi_a$, it is $\tilde{H}$ rather than $H$ which has the role of time translation generator of perturbations
\begin{equation} \label{a10}
	\delta\dot{\phi}_a({\bf x},t)=i\Big[\tilde{H}[\phi(t),\pi(t);t],\delta\phi_a({\bf x},t)\Big] \;,~~~~~~~ \delta\dot{\pi}_a({\bf x},t)=i\Big[\tilde{H}[\phi(t),\pi(t);t],\delta\pi_a({\bf x},t)\Big] \;.
\end{equation}
It follows from (\ref{a10}) that if we know perturbations in a given time $t_0$ then we are able to find them in any time through a unitary time translation 
\begin{equation}
	\delta\phi_a(t)=U^{-1}(t,t_0)\delta\phi_a(t_0)\,U(t,t_0)\;,~~~~
	\delta\pi_a(t)=U^{-1}(t,t_0)\delta\pi_a(t_0)\,U(t,t_0)\;,
\end{equation}
with a limit
\begin{equation}
	U(t_0,t_0)=1\;.
\end{equation}
Here $U(t,t_0)$ is defined by a dynamical equation
\begin{equation}\label{a11}
	\frac{d}{dt}U(t,t_0)=-i\,\tilde{H}[\delta\phi(t),\delta\pi(t);t]\,U(t,t_0).
\end{equation}
One can solve (\ref{a11}) but there is a simple way to find a more useful solution for it. We could decompose $\tilde{H}$ into a quadratic term and other high order ones
\begin{equation}
	\tilde{H}[\delta\phi(t),\delta\pi(t);t]=H_0[\delta\phi(t),\delta\pi(t);t]+H_i[\delta\phi(t),
	\delta\pi(t);t]\;.
\end{equation}
Now by using of interaction picture we define new quantities
\begin{equation} \label{a12}
	\delta\dot{\phi}^I_a(t)=i\Big[H_0[\delta\phi^I(t),\delta\pi^I(t);t],\delta\phi^I_a(t)\Big] \;,~~~~~~~ \delta\dot{\pi}^I_a(t)=i\Big[H_0[\delta\phi^I(t),\delta\pi^I(t);t],\delta\pi^I_a(t)\
\end{equation}
with initial conditions
\begin{equation}
	\delta{\phi}^I_a(t_0)=\delta{\phi}_a(t_0)\;,~~~~~
	\delta{\pi}^I_a(t_0)=\delta{\pi}_a(t_0)\;.
\end{equation}
It is worth to note that because of the quadratic form of $H_0$, the interaction picture operators behave like free fields so the time dependency of $H_0$ just comes from the explicit-time part of it. Again (\ref{a12}) can be written as a unitary transformation 
\begin{equation}
	\delta\phi_a^I(t)=U^{-1}_0(t,t_0)\delta\phi_a(t_0)U_0(t,t_0)\;,~~~~
	\delta\pi_a^I(t)=U^{-1}_0(t,t_0)\delta\pi_a(t_0)U_0(t,t_0)\;,
\end{equation}
which $U_0(t,t_0)$ satisfies the following equation
\begin{equation}\label{a13}
	\frac{d}{dt}U_0(t,t_0)=-i\,H_0[\delta\phi(t_0),\delta\pi(t_0);t]\,U_0(t,t_0)
\end{equation}
with the same initial condition which we imposed for $U(t,t_0)$ in time $t_0$. By use of (\ref{a11}) and (\ref{a13}), we have
$$
\frac{d}{dt}\Big[U_0^{-1}(t,t_0)U(t,t_0)\Big]=-iU_0^{-1}(t,t_0)H_I[\delta\phi(t_0),\delta\pi(t_0);t]U(t,t_0)\;.
$$
It is easy to see 
\begin{equation}
	U(t,t_0)=U_0(t,t_0)F(t,t_0)\;,
\end{equation}
where
\begin{equation}\label{a14}
	\frac{d}{dt}F(t,t_0)=-iH_I(t)F(t,t_0)\;,~~~~F(t_0,t_0)=1\;.
\end{equation}
and $H_I(t)$ is the interaction Hamiltonian in the interaction picture:
\begin{equation}
	H_I(t)\equiv U_0(t,t_0)H_i[\delta\phi(t_0),\delta\pi(t_0);t]U_0^{-1}(t,t_0)=H_I[\delta\phi^I(t),\delta\pi^I(t);t].
\end{equation}
By direct recursive integrating, one would be able to find solution of (\ref{a14}) as a well-known time ordering series
\begin{equation}
	F(t,t_0)=T\exp\left(-i\int_{t_0}^t H_I(t)\,dt\right).
\end{equation}
Now, we could find the time evolution of any quantity $Q(t)$ as follows
\begin{eqnarray}\label{a14}
	Q(t)  &=& F^{-1}(t,t_0)\,Q^I(t)F(t,t_0) \nonumber\\&=& \left[\bar{T}\exp\left(i\int_{t_0}^t H_I(t)\,dt\right)\right]\,Q^I(t)\,\left[T\exp\left(-i\int_{t_0}^t H_I(t)\,dt\right)\right]\;,
\end{eqnarray}
where $Q(t)$ is any $\delta\phi({\bf x},t)$ or $\delta\pi({\bf x},t)$ or any product of the $\delta\phi$s.
The important quantity which we need to compute is density matrix of wave function and then any N-point function of the $\delta\phi$s. More precisely we want to compute any quantity of the following form
\begin{equation}
	\langle \phi_{k_1} \phi_{k_2}\ldots\phi_{k_n}\rangle 
\end{equation}\\

Here, this expectation value is calculated in BD vacuum in the time $t$ after $t_0$. Note that, the time evolution of BD vacuum is given by action of $U(t,t_0)$ (in \ref{a11}) on it. The above observables could be written in Schr\"{o}dinger picture  as follows 
\begin{equation}
	\langle \phi_{k_1} \phi_{k_2}\ldots\phi_{k_n}\rangle =Tr\left[\phi_{k_1} \phi_{k_2}\ldots\phi_{k_n} \rho(t) \right]
\end{equation}
which $\rho(t)$ is computed from (\ref{a14}).

\clearpage

% Create the reference section using BibTeX:

\bibliographystyle{JHEP}
\bibliography{draft}
\end{document}